\documentclass[conference]{IEEEtran}
\IEEEoverridecommandlockouts
\usepackage[dvips]{graphicx}
  \usepackage{amsmath}
  \usepackage{amssymb}
  \usepackage{subfigure}
  \usepackage{booktabs}
  \usepackage{multirow}
  \usepackage{algpseudocode}
  \usepackage{lineno}
  \usepackage{multirow}
  \usepackage{color}
  \usepackage{url}
  \usepackage{diagbox}
  \usepackage{relsize}
  \usepackage{bm}
  \usepackage[misc]{ifsym}

  \usepackage[compress]{cite}

  \usepackage{algorithm}
  \usepackage{algorithmicx}
  \usepackage{algpseudocode}
  \floatname{algorithm}{Algorithm}

  \graphicspath{{./}}

\def\BibTeX{{\rm B\kern-.05em{\sc i\kern-.025em b}\kern-.08em
    T\kern-.1667em\lower.7ex\hbox{E}\kern-.125emX}}
\begin{document}

\title{Algorithm and Architecture for Path Metric Aided Bit-Flipping Decoding of Polar Codes\\
\thanks{This work was supported in part by the NSF of China (Grant No. 61874140).}
}

\author{\IEEEauthorblockN{Yu Wang, Lirui Chen, Qinglin Wang, Yang Zhang, Zuocheng Xing}
\IEEEauthorblockA{National Laboratory for Parallel and Distributed Processing,
National University of Defense Technology,
Changsha, China \\
Email: \{wangyu16,chenlirui14,wangqinglin,zhangyang,zcxing\}@nudt.edu.cn}
}

\maketitle

\begin{abstract}
    Polar codes attract more and more attention of researchers in recent years, since its capacity achieving property. However, their
    error-correction performance under successive cancellation (SC) decoding is inferior to other modern channel codes at short or moderate blocklengths. SC-Flip (SCF) decoding algorithm
    shows higher performance than SC decoding by identifying possibly erroneous decisions made in initial SC decoding and flipping them in the sequential decoding attempts.
    However, it performs not well when there are more than one erroneous decisions in a codeword. In this paper, we propose a path metric aided bit-flipping decoding algorithm to identify and 
    correct more errors efficiently. In this algorithm, the bit-flipping list is generated based on both log likelihood ratio (LLR) based path metric and bit-flipping metric. The path metric is 
    used to verify the effectiveness of bit-flipping. In order to reduce the decoding latency and computational complexity, its corresponding pipeline architecture is designed.
    By applying these decoding algorithm and pipeline architecture, an improvement on error-correction performance can be got up to 0.25dB compared with SCF decoding at frame error rate of $10^{-4}$, with low average decoding latency. 
\end{abstract}


\section{Introduction}
Polar codes \cite{Arikan2008Channel} are the first channel codes proven to achieve the capacity of various communication channels and have been selected for the control channel in the 5G enhanced Mobile
  BroadBand (eMBB) scenario \cite{3GPPstandard}. However, for short to moderate blocklengths,
  the performance of successive cancellation (SC) decoding is worse than that of Turbo codes or low-density parity-check (LDPC) codes. To overcome this limitation, SC 
  list (SCL) decoding \cite{Tal2012List} was introduced to improve the performance at the cost of increased computational 
  complexity and decoding latency.

  In recent researches, the successive cancellation flip (SCF) decoding proposed in \cite{Afisiadis2014A} was shown to be capable of providing error-correction performance close to that of 
  SCL decoding with a small list size, while keeping the computational complexity close to that of SC. The idea of SCF decoder is to allow multiple subsequent decoding attempts to opportunistically
  correct the erroneous decision made in initial SC decoding by flipping the most unreliable bit. Modifications to SCF decoding are proposed in \cite{Giard2017Fast,Condo2018Improved,Furkan2018Bit-Flipping} to reduce the decoding latency and implementation
  complexity. However, these decoding methods focus on correcting the first error and can not identify more than one erroneous decisions, which limits their error-correction performance.

  In order to enhance the performance of SCF decoding, several improvements have been proposed to correct more erroneous decisions \cite{Furkan2017Partitioned,Zhang2017Progressive,Fazeli2017Viterbi,Chandesris2017An,Chandesris2018Dynamic}.
  In \cite{Furkan2017Partitioned}, they subdivide the codeword into several partitions, on which SCF is run individually. However, by adopting their method to correct more errors, 
  the erroneous bits need to distribute just in different partitions evenly, which limits its correcting capability. In \cite{Zhang2017Progressive}, they investigate the distribution of the first erroneous
  bit and restrict the search scope of flipping bits to a subset of information bits. By iteratively modifying the subset, their method can identify multiple incorrect bits. However, the reason of 
  erroneous decisions is not only due to the transmitting capability of the subchannel itself, but also the condition of current channel noise. So we can not determine the positions of flipping bits by
  only considering the codeword itself.
  
  The dynamic SCF (DSCF) decoding algorithm proposed in \cite{Chandesris2018Dynamic} shows a promising way to identify multiple erroneous decisions, while their method is not efficient to correct them.
  Different from their work, in our method, we generate the bit-flipping list based on both path metric and bit-flipping metric. The path metric of each decoding attempt can be used as a feedback to verify the effectiveness of bit-flipping attempt.
  Basing on effective bit-flipping attempt, we could identify multiple erroneous decisions step by step.
  In order to reduce the decoding latency and implementation complexity, a pipeline decoding architecture is designed. 

  The remainder of this work is organized as follows: in Section II, an overview of polar codes, SCF decoding, and DSCF decoding is presented. In Section III, the proposed decoding algorithm
  and its corresponding pipeline architecture are detailed. Section IV reports the simulation results, and then conclusions are drawn in Section V. 

  \section{preliminary}

  \subsection{Polar Codes}
  Polar codes characterized by $(N,K,\mathcal{I})$ can achieve channel capacity via the phenomenon of channel polarization \cite{Arikan2008Channel}. The channel polarization
  theorem states that, as the blocklength $\textit{N}$ goes to infinity, a polarized subchannel becomes either a noiseless channel or a pure noise channel. By transmitting information bits over the reliable
  subchannels and transmitting frozen bits which are known by both transmitter and receiver over the unreliable subchannels, polar codes can achieve the channel capacity. Hence, constructing
  a polar code is equivalent to find the $\textit{K}$ most reliable subchannels over which the information bits are transmitted, with a set $\mathcal{I}$ indicating  
  the locations of these subchannels. Many construction methods have been proposed to calculate the reliability of subchannels. In our work, we use
  the Gaussian approximation (GA) based density evolution method proposed in \cite{Wu2014Construction}, since it is popular in the construction of polar codes for its good tradeoff between the 
  complexity and performance.
  
  The encoding process of a polar code can be represented with a matrix multiplication like: 
  \begin{align}\label{PASCF_EQ1}
    x = u{G_N},\;{\rm{where}}\;{G_N} = B{\left[ {\begin{array}{*{20}{c}}
      1 & 0  \\
      1 & 1  \\
   \end{array}} \right]^{ \otimes n}}
  \end{align}
  The vector $\textbf{u}$ hold the information bits denotes the source codeword to be encoded, vector $\textbf{x}$ denotes the encoded codeword and ${G_{N}}$ is the generator matrix, while $\otimes$ denotes the Kronecker product,
  $\textit{B}$ is a bit-reversal permutation matrix. 
  
  As for the decoding, we denote by $\textbf{y}$ the data received from the channel and use them as the decoder
  inputs. The decoder's output is denoted by $\hat{u}^{N}_1$, where $\hat{u}_i$ is the estimate of the bit $u_i$ by hard decision.
  This hard decision is made according to the log likelihood ratio (LLR)
  $L_i=log(\frac{Pr(\rm{y},\hat{u}_{1}^{i-1})\mid u_{i}=0}{Pr(\rm{y},\hat{u}_{1}^{i-1})\mid u_{i}=1})$ by using the hard decision function $\textit{h}$:
  \begin{align}\label{PASCF_EQ2}
    {\hat u_i} = h({L_i}) = \left\{ {\begin{array}{ll}
      {u_i}\;&{\rm{if}}\;i \notin \mathcal{I}  \\
      \frac{{1 - {\mathop{\rm sign}\nolimits} ({L_i})}}{2}\;&{\rm{if}}\;i \in \mathcal{I}  \\
   \end{array}} \right. 
  \end{align}
  where $\mathop{\rm sign}(L_i) = \pm1$. At the same time, the LLRs at different calculation stage \textit{l} are computed iteratively by follows:
  \begin{equation}\label{PASCF_EQ3}
    L_{l,i}= \left\{
   \begin{array}
   {l@{\quad \quad}l}
   f(L_{l+1,i};L_{l+1,i+2^{l}}) & \text{if } \frac{i}{2^{l}}\text{is even} \\
   g(\hat{s}_{l,i-2^{l}};L_{l+1,i-2^{l}};L_{l+1,i}) & \text{otherwise} \\
   \end{array}
   \right.
  \end{equation}
  where $\hat{s}$ denotes the partial sum of $\hat{u}^{i-1}_{1}$. And in the LLR domain, the function $\textit{f}$ and $\textit{g}$ perform the following calculation for given inputs LLRs $L_a$ and $L_b$:
  
  \begin{align}
    \label{PASCF_EQ4}
    f({L_a},{L_b})&=\log (\frac{{{e^{{L_a} + {L_b}}} + 1}}{{{e^{{L_a}}} + {e^{{L_b}}}}}) \\
    \label{PASCF_EQ5}
    g({L_a},{L_b},{u_s})&={( - 1)^{{u_s}}}{L_a} + {L_b}
  \end{align}
  
  \subsection{Successive Cancellation Flip Decoding}
  The SCF decoding is a slightly-modified SC decoding algorithm, characterized by a number of extra decoding attempts, where 
  several unreliable bits are flipped from its initial SC decoding. The decoding procedure of SCF is that, after the 
  first SC decoding pass, the concatenated cyclic redundancy check (CRC) is verified. In case it matches, the decoding procedure stops and the estimated $\hat{u}^{N}_1$
  is output. Otherwise, a list of positions of the least reliable estimated bits is built and then another SC decoding pass is launched. In this pass, once the location of the information bit that corresponds to the least reliable
  bit is reached, that estimated bit is flipped before subsequent SC decoding. Once an SC decoding pass has finished, the CRC is verified again. 
  This procedure is repeated until the CRC pass or a predetermined maximum number $\textit{T}$ of decoding attempts is reached.
  However, since the concatenated CRC could not indicate the number of erroneous decisions and their positions, the performance of SCF decoding is limited by a hypothetical decoder, called SC-oracle decoder \cite{Afisiadis2014A}, which can accurately avoid
  all first wrong decisions.
  
  \subsection{Dynamic SCF Decoding}
  The DSCF decoding aimed to correct multiple erroneous bits was proposed in \cite{Chandesris2018Dynamic}. This decoding method is characterized by a bit-flipping list $\mathcal{L}_{flip}$, 
  which updates after
  every decoding attempt. It contains $\textit{T}$ bit-flipping sets $\{ {\mathcal{E} _1}, \cdots ,{\mathcal{E} _{{\omega}}}\}$ with the highest probability to correct the trajectory of the SC decoding.
  The $\omega$ denotes the maximum number of bits that can be corrected by this list. They name it the noise order, which indicates the correcting capability of the bit-flipping list. By adopting this definition,
  the order of SCF decoding is order-1. 
  
  The DSCF decoding algorithm builds the bit-flipping list by using a new bit-flipping metric $M_\alpha$ \cite{Chandesris2017An}, which takes into account the serial nature of the SC decoder. This metric has a much higher 
  probability to find the first error that occurred during the sequential decoding process than the absolute value of LLRs.
  The method of calculating the probability of a flip set $\mathcal{E}_{\omega}$ with $\omega$ flipping bits to correct the trajectory of SC is close to that used to calculate
  $P({\mathcal{C}_i})=P({\hat u_i} \ne {u_i}|\hat u_1^{i - 1} = u_1^{i - 1})$ in \cite{Wu2014Construction}.
  By using this method, the probability $P(\mathcal{E}_{\omega})$ can be computed by the following expression:
  
  \begin{align}
    \label{PASCF_EQ6}
    P(\mathcal{E}_{\omega}) = \prod\limits_{j \in {\mathcal{E} _\omega }} {{p_e}(\hat u{{[{\mathcal{E} _{\omega  - 1}}]}_j}) \cdot \prod\limits_{\scriptstyle j < {i_\omega } \atop 
    \scriptstyle j \in \mathcal{I}\backslash {\mathcal{E} _\omega }} {(1 - {p_e}(\hat u{{[{\mathcal{E} _{\omega  - 1}}]}_j}))} }
  \end{align}
  
  However, since the computation of ${p_e}(\hat u{{[{\mathcal{E} _{\omega  - 1}}]}_j})$ is a hard task, it can be approximately replaced by ${q_e}(\hat u{{[{\mathcal{E} _{\omega  - 1}}]}_j})= {1 \mathord{\left/
  {\vphantom {1 {(1 + \exp (|{\rm{L}}{{[{\mathcal{E} _{\omega  - 1}}]}_j}|))}}} \right.
  \kern-\nulldelimiterspace} {(1 + \exp (|{\rm{L}}{{[{\mathcal{E} _{\omega  - 1}}]}_j}|))}}$. By this way,
  they defined their bit-flipping metric as:
  \multlinegap = 0pt
  \begin{multline}
    \label{PASCF_EQ7}
    {M_\alpha }({\mathcal{E} _\omega }) = \prod\limits_{j \in {\mathcal{E} _\omega }} {\left( {\frac{1}{{1 + \exp (\alpha |{\rm{L}}{{[{\mathcal{E} _{\omega  - 1}}]}_j}|)}}} \right)}  \cdot \\
    \prod\limits_{\scriptstyle j < {i_\omega } \atop 
    \scriptstyle j \in \mathcal{I}\backslash {\mathcal{E} _\omega }} {\left( {\frac{1}{{1 + \exp ( - \alpha |{\rm{L}}{{[{\mathcal{E} _{\omega  - 1}}]}_j}|)}}} \right)}
  \end{multline}
  Specially for the initial SC decoding pass, the ${M_\alpha }(i)$ of each information bit can be calculated as:
  \begin{equation}
    {M_\alpha }(i) = \frac{1}{{1 + \exp (\alpha |{{\rm{L}}_i}|)}} \cdot \prod\limits_{\scriptstyle j < i \atop 
    \scriptstyle j \in \mathcal{I}} {\left( {\frac{1}{{1 + \exp ( - \alpha |{{\rm{L}}_j}|)}}} \right)}
  \end{equation}
  
  The difference of procedure between SCF decoding and DSCF decoding lies in the updating of the bit-flipping list. For DSCF decoding, after each decoding attempt, new flipping bit would be 
  added to the current flip set and its corresponding ${M_\alpha }$ would be computed. If the ${M_\alpha }$ is greater than the least one in the list, the bit-flipping set will be inserted to 
  the list.
  
  \section{Improved Successive Cancellation Flip Decoding}
  In this section, we propose a path metric aided bit-flipping decoding to improve the performance of SCF decoder to correct multiple erroneous decisions. And its corresponding pipeline 
  architecture is designed to reduce decoding latency.
  
  \subsection{Path Metric Aided SCF Decoding Algorithm}
  In comparison of different bit-flipping decoding algorithms proposed in current researches, we find that for many wrong estimated codewords the first error is always in the initial bit-flipping list. 
  However, the correct codeword can not be got at the last since that there are more than one errors caused by
  channel noise in a codeword or that the decoding algorithm can not find out all of the errors in the limited attempts. 
  Based on this view, we make simulations to evaluate the correct ratio when the first error bit is in the initial bit-flipping list built by using the bit-flipping metric proposed in \cite{Chandesris2018Dynamic}.
  
  \begin{figure}[b]
    \centering
    \includegraphics[width=8cm]{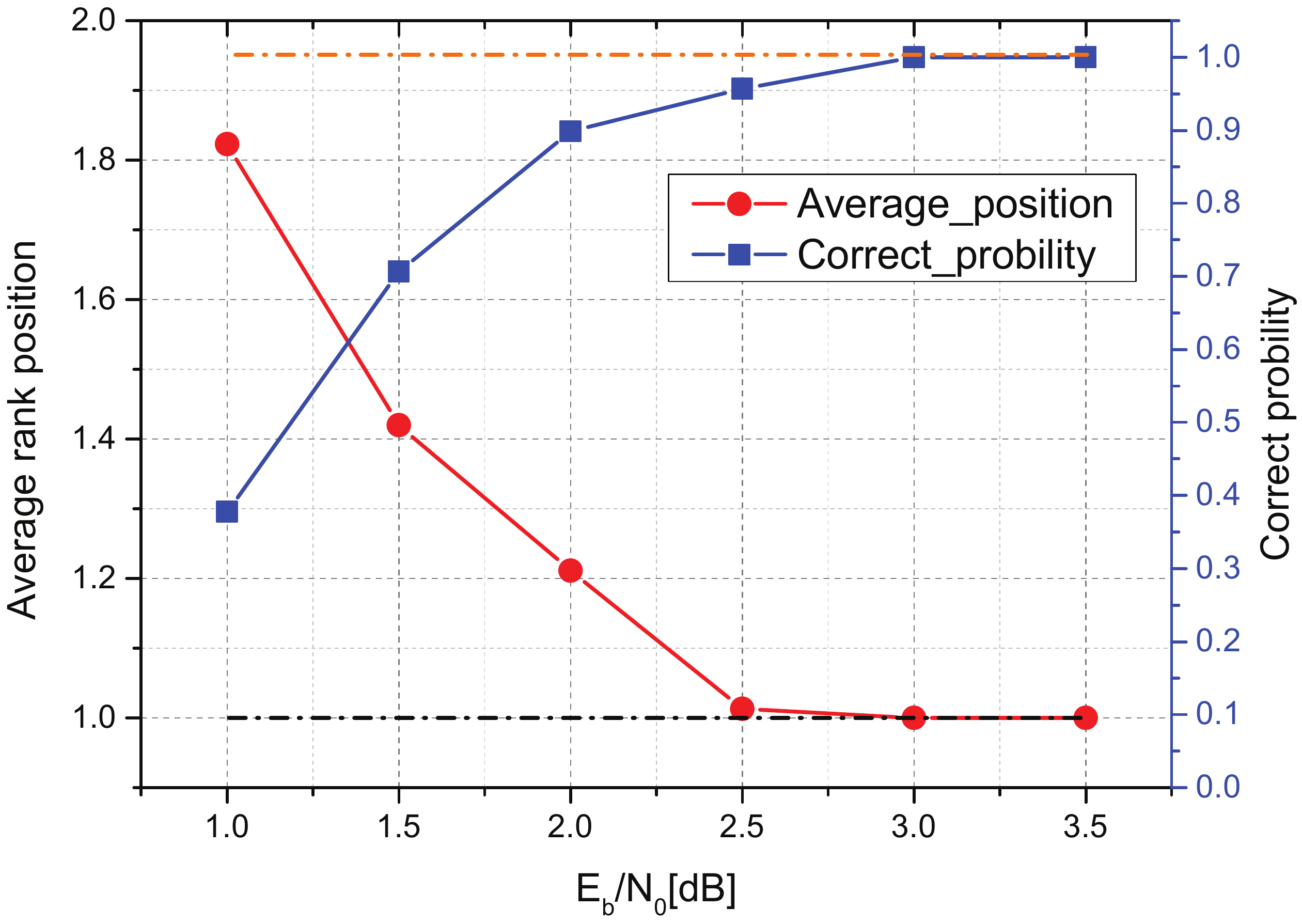}
    \caption{The correct ratio of DSCF when the first error bit is in the bit-flipping list and the average rank of the first error bit in 
    the bit-flipping list.}\label{Fig.aver_correct}
  \end{figure}
  
  From Fig.\ref{Fig.aver_correct}, we can observe that the correct ratio is unsatisfactory at low ${{{E_b}} \mathord{\left/
  {\vphantom {{{E_b}} {{N_0}}}} \right.
  \kern-\nulldelimiterspace} {{N_0}}}$ regime and that the performance of DSCF decoding is dependent on its ranking in the initial bit-flipping list. 
  This means that the bit-flipping set at the top of the list has more probability to try more than one bit-flipping positions in the limited decoding attempts.
  When the bit-flipping set containing the accurate bit-flipping bits lies at the bottom of the list, it may be taken out of the list 
  in the subsequent list updating or could not get enough attempts to find all of the errors before reach the predetermined $T$.
  
  In the above simulations, we also found that if the first error position is in the initial bit-flipping list, its corresponding codeword almost has the smallest LLR-based path metric \cite{Balatsoukas2014LLR}. As shown in Fig.\ref{Fig.aver_correct}, 
  the set containing the accurate bit-flipping bit ranks top in the list sorted by LLR-based path metric. Hence, we introduce the
  LLR-based path metric as a feedback to the bit-flipping list generating.
  
  The procedure of our Path Metric Aided SCF (PMA-SCF) decoding algorithm is that: after the first SC-decoding pass, the initial bit-flipping
  list is built based on the bit-flipping $M_\alpha$ value, and then the bit-flipping sets will be attempted one by one until the CRC matches or all bit-flipping sets have been attempted.
  During these decoding attempts, the LLR-based path metric of each attempt will be calculated, while a new bit-flipping list with
  order-2 will be built with its corresponding former bit-flipping set. This list will be sorted first by the path metric and then by the $M_\alpha$ value.
  That is to say, whether it has the priority to be a start point for correcting multiple errors is determined by the path metric,
  while whether a bit should be added to the bit-flipping set is determined by the $M_\alpha$ value. Then new decoding attempt launches according to this new list. 
  The detail of this decoding algorithm is described in Algorithm.\ref{alg:PMA-SCF}, \ref{alg:bitflipdecode} and \ref{alg:genlist}.
  
  \begin{algorithm}[h]
    \caption{Path Metric Aided SCF Decoding Algorithm}
    \label{alg:PMA-SCF}
    \begin{algorithmic}[1]
      \Procedure {PMA-SCF}{$y^N_1,T,\mathcal{I}$}
      \State $(\hat u_1^N,L(y_1^N,\hat u_1^{i - 1}|{u_i}),PM_{init}) \leftarrow {\rm{SC}}(y_1^N,\mathcal{I},\varnothing)$
      \If{$T > 1$ and CRC($\hat u_1^N$) = failure}
      \State $(\mathcal{L}_{init},\mathcal{M}_{init}) \leftarrow$ Init$({L_{i \in \mathcal{I}}(y_1^N,\hat u_1^{i - 1}|{u_i})},T)$
      \State $(\mathcal{L},\mathcal{M},\mathcal{P})\leftarrow(\mathcal{L}_{init},\mathcal{M}_{init},PM_{init})$
      \While{$\mathcal{L} \ne \varnothing$}
      \State $(\mathcal{L},\mathcal{M},\mathcal{P})\leftarrow$BitFlip\_Decode$(y_1^N,\mathcal{I},\mathcal{L},\mathcal{M},\mathcal{P})$
      \EndWhile
      \Else{ return $\hat{u}^N_1$} 
      \EndIf
      \EndProcedure
    \end{algorithmic}
    \end{algorithm}
  
    \begin{algorithm}[h]
      \caption{Bit-flipping Decoding}
      \label{alg:bitflipdecode}
      \begin{algorithmic}[1]
        \Procedure{BitFlip\_Decode}{$y_1^N,\mathcal{I},\mathcal{L},\mathcal{M},\mathcal{P}$}
        \State $T\leftarrow$size\_of$(\mathcal{L})$
        \For{$j \leftarrow 1$ \textbf{to} $T$}
        \State $(\hat u_1^N,\{ L{[{\mathcal{E}_j}]_i}\},PM_j) \leftarrow {\rm{SC}}(y_1^N,\mathcal{I},\mathcal{E}_j \in \mathcal{L})$
        \If {CRC($\hat u_1^N$) = success} {return $\hat{u}^N_1$}
        \Else
        \If {$PM_{j}>\mathcal{P}_m$} {$(\mathcal{L}_j,\mathcal{M}_j) \leftarrow \varnothing$}
        \Else {$(\mathcal{L}_j,\mathcal{M}_j) \leftarrow$ Gen$(\{ L{[{\mathcal{E}_j}]_i}\},\mathcal{E}_j,\mathcal{M}_{least})$}
        \EndIf
        \EndIf
        \EndFor
        \State $(\mathcal{L},\mathcal{M}) \leftarrow$ Sort$(\mathcal{L}^{T}_{1},\mathcal{M}^{T}_{1},PM^{T}_{1})$
        \State return $(\mathcal{L},\mathcal{M},PM^{T}_{1})$
        \EndProcedure
      \end{algorithmic}
    \end{algorithm}
  
    \begin{algorithm}[h]
      \caption{Generate the bit-flipping list}
      \label{alg:genlist}
      \begin{algorithmic}[1]
        \Procedure {Gen}{$\{ L{[{\mathcal{E}_j}]_i}\} _{i \in \mathcal{I}},\mathcal{E}_j,\mathcal{M}_{least}$}
        \For{$i = {\rm{last}}({\mathcal{E}_j}) + 1, \ldots ,N$ and $i \in \mathcal{I}$} 
        \State $\mathcal{E} = {\mathcal{E}_j} \cup \{ i\} ;m = {M_\alpha }(\mathcal{E})$
        \If{$m > {\mathcal{M}_{least}}$} {Insert\_list$(\mathcal{L},\mathcal{M},\mathcal{E},m)$}
        \EndIf
        \EndFor
        \State return ($\mathcal{L},\mathcal{M}$)
        \EndProcedure
      \end{algorithmic}
    \end{algorithm}
  
    \begin{figure*}[t]
      \centering
      \includegraphics[width=16cm]{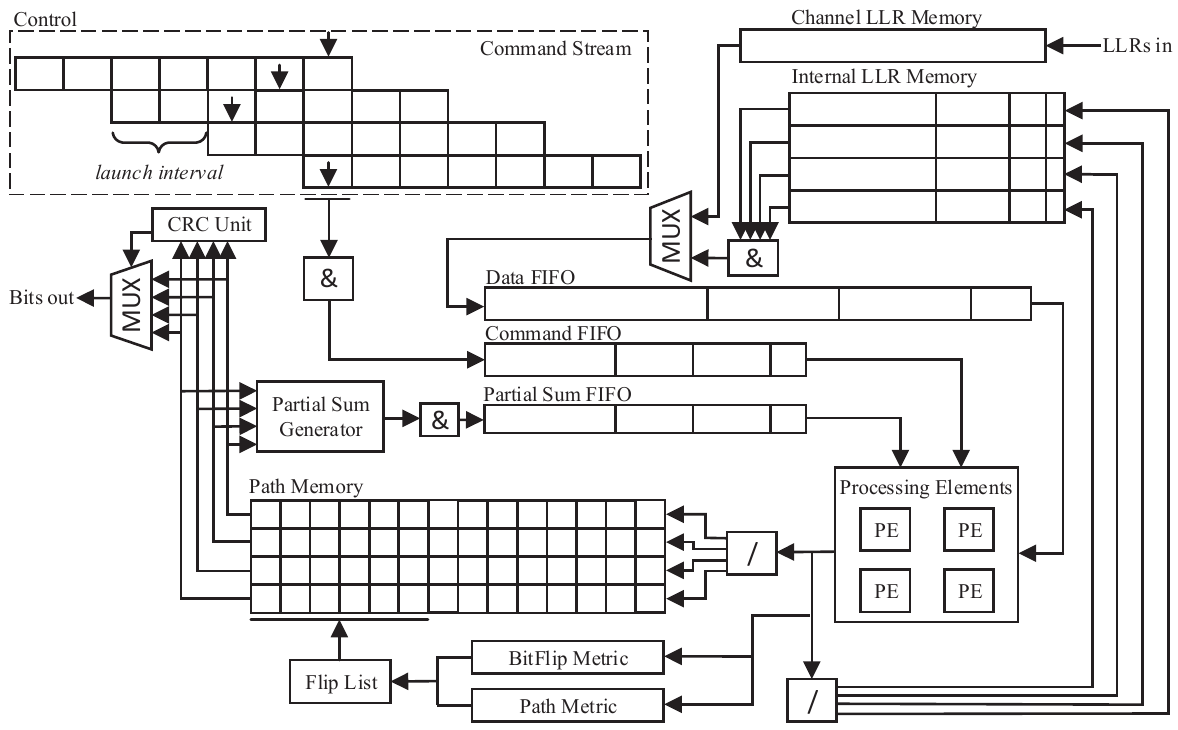}
      \caption{The pipeline architecture of our proposed decoding algorithm with four parallel bit-flipping decoding attempts.}\label{Fig.pipeline}
    \end{figure*}
  
    In Algorithm.\ref{alg:bitflipdecode}, $\mathcal{P}_m$ denotes the path metric of the decoding pass, whose bit-flipping set current decoding attempt extends.
    The function Sort firstly sorts the $\mathcal{L}_j$ by its $PM_{j}$, and then sorts the bit-flipping set $\mathcal{E}$ in list $\mathcal{L}_j$ by its ${M_\alpha }(\mathcal{E})$.
    All these sorted $\mathcal{L}_j$ constitute the new bit-flipping list $\mathcal{L}$.
  
  \subsection{Pipeline Architecture for Bit-flipping Decoding}
  By using the path metric as the feedback, it is inevitable to increase the decoding latency if we still use the decoding
  architecture of SC. In order to reduce the decoding latency, we design a pipeline architecture to realize parallel decoding of different attempts.
  Different from the parallel architecture adopted by SCL decoder, our pipeline architecture does not need too many processing elements to calculate 
  the LLR data of different decoding attempts at the same time, since there is no data dependency between different attempts.
  As a result, the usage ratio of processing elements in our pipeline is much higher than that of SCL decoder.
  
  The pipeline decoding is realized by splitting the command stream from its corresponding data. Since the different decoding attempts
  have the same decoding schedule, we can use only one set of command with several pointers to realize the control of different decoding attempts. 
  As shown in Fig.\ref{Fig.pipeline}, there are four pointers indicating the current decoding stage of its corresponding decoding attempt. Each command in the command stream contains
  the decoding stage information and the type of decoding function ($f$ or $g$).
  
  According to the pointers, the commands are fetched to the command FIFO, while the corresponding data of different attempts are
  fetched to the data FIFO. Then the processing elements controlled by the current commands work to process the data in the data FIFO.
  Each processing element could execute the $f$ or $g$ calculation and hard decision function $h$ according to the stage and function type. Since the data buffered in data FIFO may not be 
  processed completely in one cycle of calculation, the remaining data will stay in the FIFO. Meanwhile, new data will be pushed into the FIFO when the data of former command have been processed, 
  which will lead to a high peak-to-average ratio of the usage of data FIFO. In order to reduce the size and peak-to-average ratio of data FIFO, launch 
  intervals are arranged among the sequential decoding attempts to avoid data with large scale to be pushed into data FIFO at the same time.
  
  The calculation results of the processing elements are sent to corresponding internal LLR memory, while the hard decision results are sent to
  the path memory. Based on the hard decisions, the partial sums are calculated. Then they are fetched to the partial sum FIFO. An insertion sorter is adopted to calculate and sort the bit-flipping metric of each bit.
  Meanwhile, the path metrics and the CRC check are computed based on the hard decisions on-the-fly. Based on the bit-flip metrics and path metrics, the flip list is generated. 
  
  Besides, by applying the latency saving technique proposed in \cite{Giard2017PolarBear}, the decoding latency 
  can be farther reduced, since the different decoding attempts have the same start point in the decoding command stream. Due to page 
  limitations, the trade-off among decoding latency, usage ratio of processing elements and memory requirements are omitted here. 
  A more detailed presentation about this architecture will be given in the full version of this paper.
  
  \section{Simulation Results}
   In this section, the frame error rate (FER) performance and the decoding latency of the 
   proposed PMA-SCF decoding algorithm are evaluated via Monte-Carlo simulations. We make simulations based on the AFF3CT \cite{aff3ct} software, which is extended with our designed decoding algorithm.
   Specially, the transmissions are run on binary phase-shift keying (BPSK) modulation and additive white Gaussian noise (AWGN) channel. All polar codes are constructed targeting an ${{{E_b}} \mathord{\left/
   {\vphantom {{{E_b}} {{N_0}}}} \right.
   \kern-\nulldelimiterspace} {{N_0}}}$
   of 3.0dB. And all CRC-aided polar codes are concatenated with a 16-CRC with generator polynomial $g(D)\; = \;{D^{16}} + {D^{12}} + {D^5} + 1$.
   In this regard, the coding rate of these polar codes is ${R} = {{(K + 16)} \mathord{\left/
   {\vphantom {{(K + 16)} N}} \right.
   \kern-\nulldelimiterspace} N}$.
  
   \begin{figure}[b]
    \centering
    \includegraphics[width=8cm]{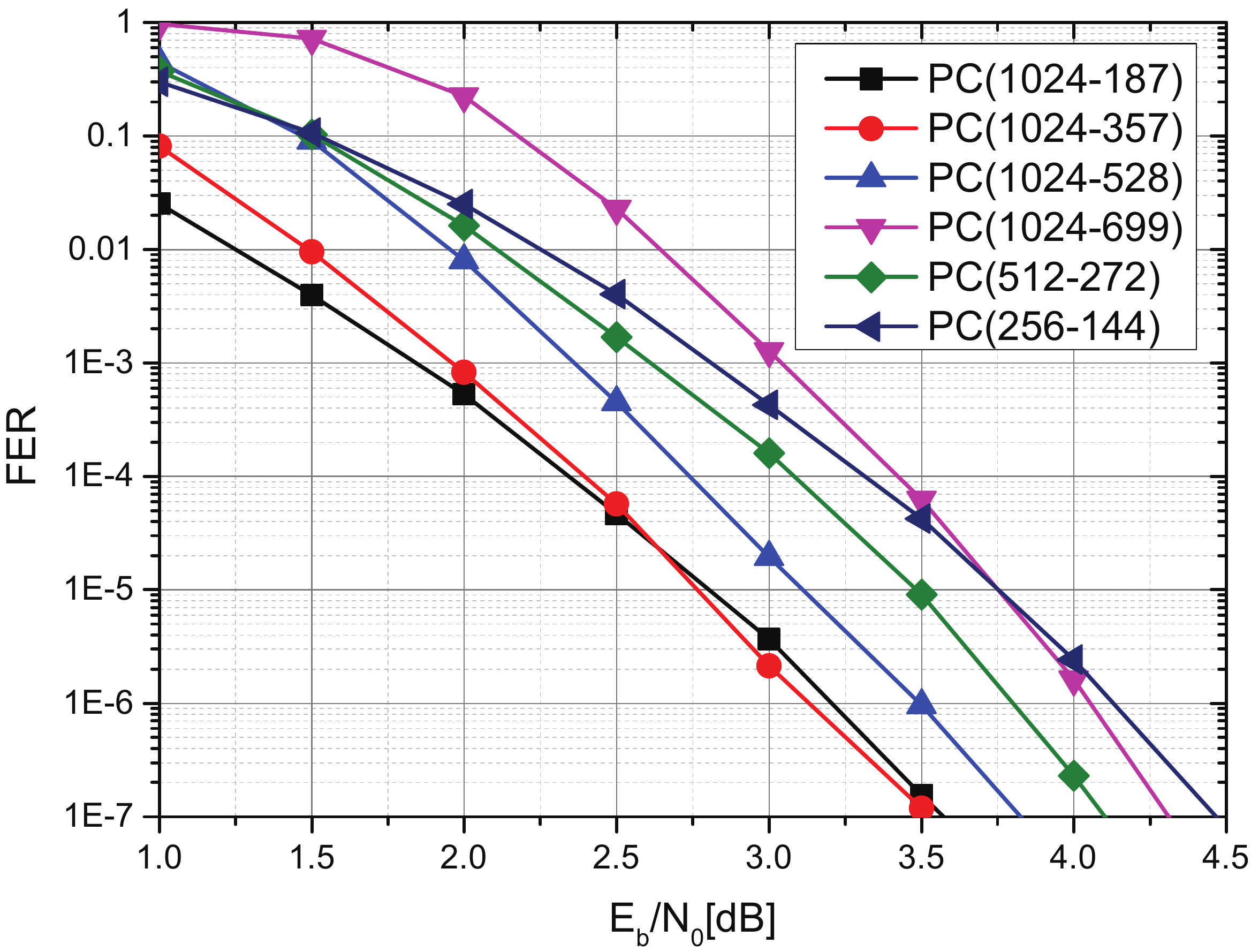}
    \caption{FER performance comparison of our proposed decoding algorithm for polar codes of length $N\in\{ 256,512,1024\}$ and $R \in \{ 1/6,1/3,1/2,2/3\}$, with predetermined maximum attempts $T=10$.}\label{Fig.coderate_compare}
  \end{figure}
  
  In Fig.\ref{Fig.coderate_compare}, we compare the performance of PMA-SCF decoder for polar codes with different blocklength and rate.
  One can observe that the polar codes with low code rate have much better performance than that with high code rate, specially at the low ${{{E_b}} \mathord{\left/
  {\vphantom {{{E_b}} {{N_0}}}} \right.
  \kern-\nulldelimiterspace} {{N_0}}}$ regime.
  The performance gap between different code rate narrows at the high ${{{E_b}} \mathord{\left/
  {\vphantom {{{E_b}} {{N_0}}}} \right.
  \kern-\nulldelimiterspace} {{N_0}}}$ regime, which demonstrates the effectiveness of the proposed decoding algorithm to correct erroneous decisions. However, the performance gap between the $R=1/2$ curve
  and the $R=2/3$ curve does not narrow much, since there are too many erroneous bits in a codeword of code rate $R=2/3$, which beyond the correction capability of PMA-SCF decoder.
  Besides, the error-correction performance increases as the blocklength increases. It can be also observed that the gaps between different blocklengths narrow quickly at high ${{{E_b}} \mathord{\left/
  {\vphantom {{{E_b}} {{N_0}}}} \right.
  \kern-\nulldelimiterspace} {{N_0}}}$ regime, since more wrong estimated
  codewords can be corrected by PMA-SCF decoding. 
  
  \begin{figure}[tb]
    \centering
    \includegraphics[width=8cm]{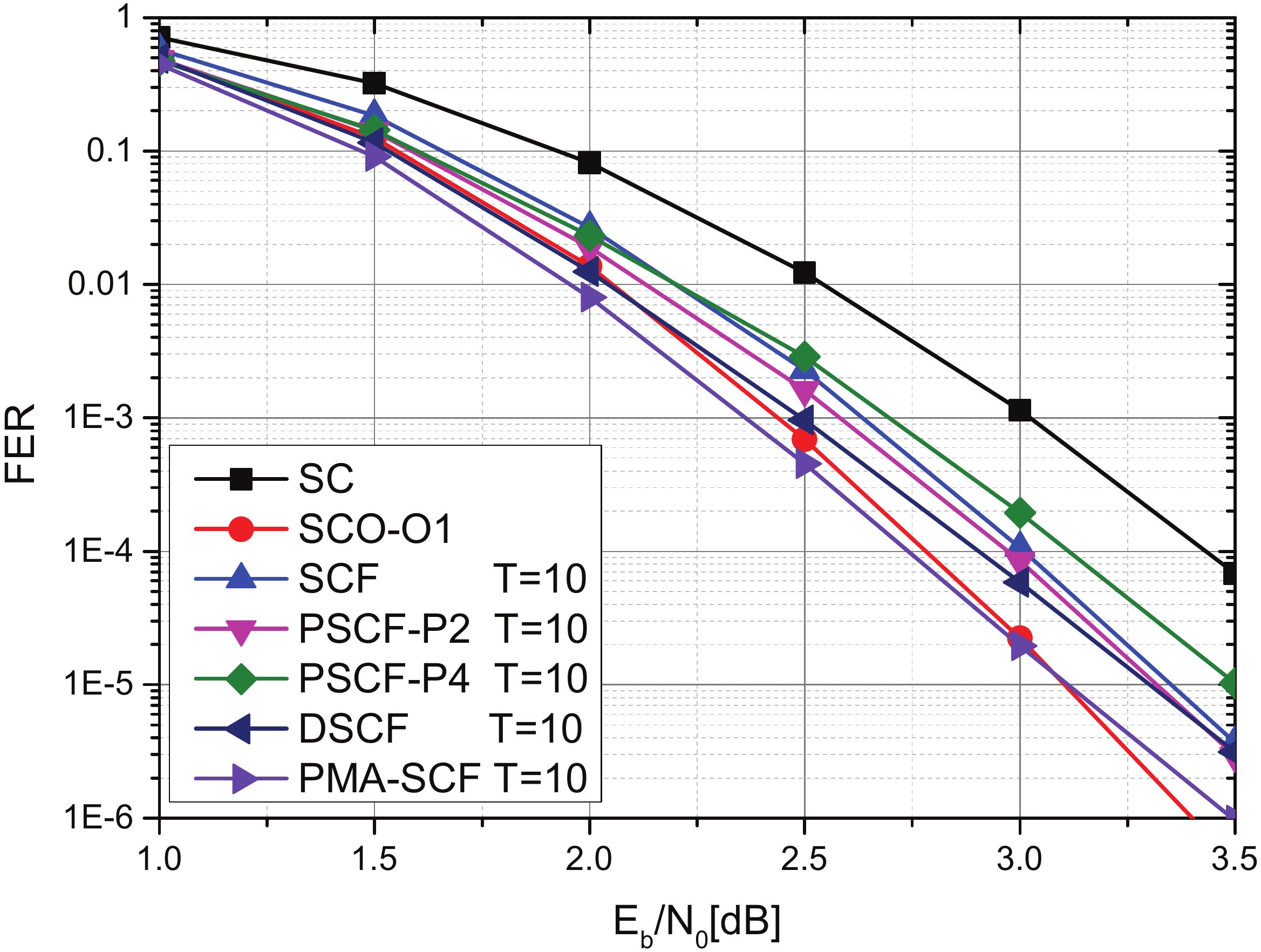}
    \caption{FER performance of our proposed decoding algorithm for polar code PC(1024-512), compared with other bit-flipping methods and oracle-assisted SC-oracle decoder with order 1 (SCO-O1).}\label{Fig.flip_compare}
  \end{figure}
  
   Fig.\ref{Fig.flip_compare} depicts the FER performance of our proposed decoding algorithm for polar code PC(1024-528) against other bit-flipping decoding algorithms, where PSCF-P2 denotes the partitioned SCF decoding \cite{Furkan2017Partitioned} with 
   divided partition P=2. In order to keep the same code rate, the concatenated CRC code for PSCF-P2 is 8-CRC, while 4-CRC for PSCF-P4. In Fig.\ref{Fig.flip_compare}, all incarnations of SCF decoding have the same predetermined maximum attempts $T=10$. 
   The FER performance of oracle-assisted SC-oracle decoder (SCO-O1) and SC decoder are used as the baseline for comparison, where SCO-O$k$ means that it can always correct the first $k$ erroneous decisions met by
   SC decoder, but no more errors can be corrected. It can be observed that our PMA-SCF decoding algorithm performs slightly better than SCO-O1 decoder in almost all cases, since its ability to correct higher-order 
   errors. However, the performance gap between PMA-SCF and SCO-O1 curves narrows at high ${{{E_b}} \mathord{\left/
   {\vphantom {{{E_b}} {{N_0}}}} \right.
   \kern-\nulldelimiterspace} {{N_0}}}$ regime and SCO-O1 outperforms PMA-SCF at ${{{E_b}} \mathord{\left/
   {\vphantom {{{E_b}} {{N_0}}}} \right.
   \kern-\nulldelimiterspace} {{N_0}}}=3.5dB$. This is due to the fact that the number of multiple errors in one codeword
   decreases as the ${{{E_b}} \mathord{\left/
   {\vphantom {{{E_b}} {{N_0}}}} \right.
   \kern-\nulldelimiterspace} {{N_0}}}$ increases and PMA-SCF could not accurately identify all first errors.
   Besides, one can observe that our proposed decoding algorithm outperforms the SCF decoding 0.25dB at FER of $10^{-4}$ with the same $T$ value and performs better than DSCF decoding at all ${{{E_b}} \mathord{\left/
   {\vphantom {{{E_b}} {{N_0}}}} \right.
   \kern-\nulldelimiterspace} {{N_0}}}$ conditions.
  
   \begin{figure}[tb]
    \centering
    \includegraphics[width=8cm]{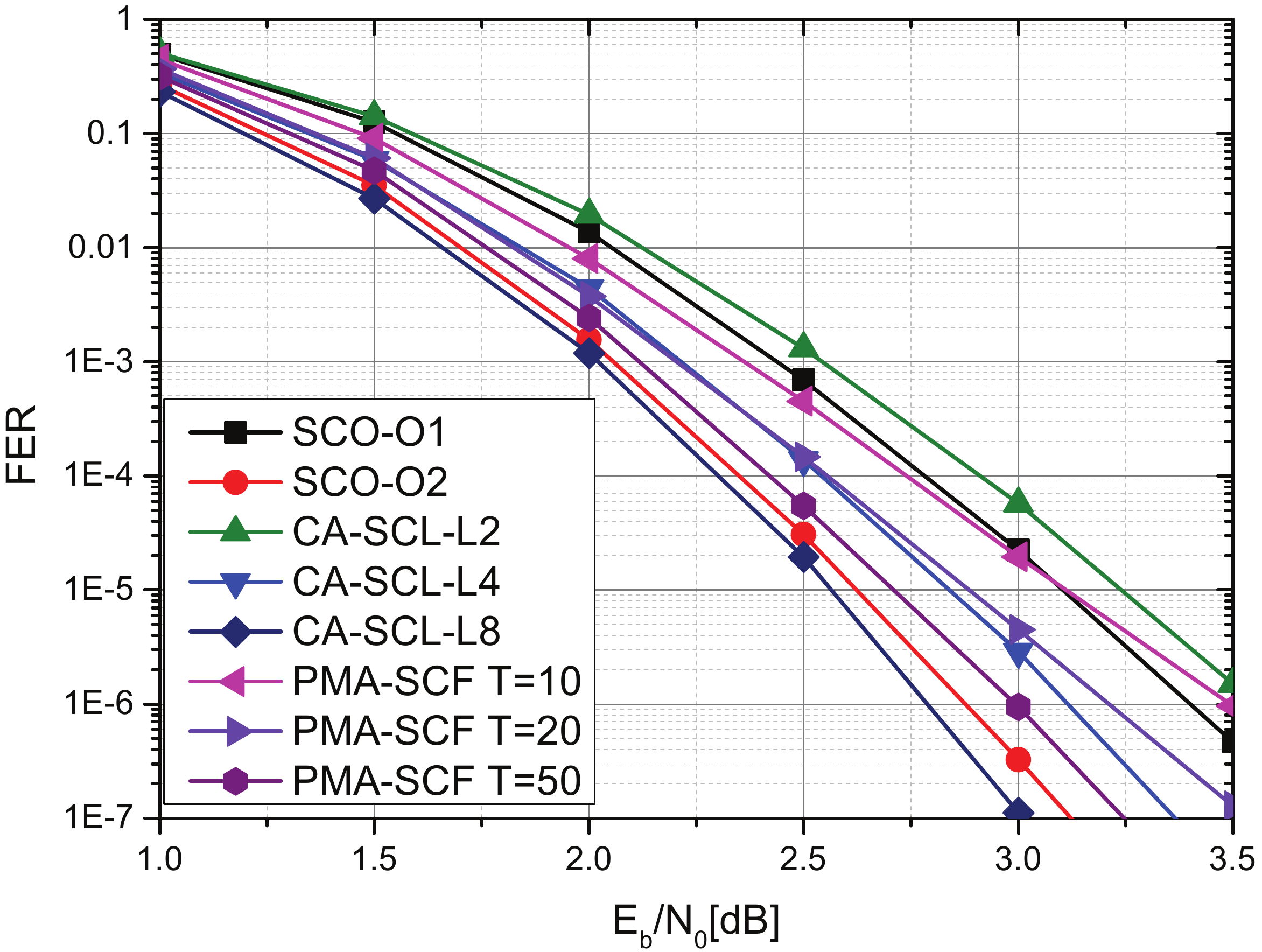}
    \caption{FER performance of PMA-SCF decoding algorithm with different predetermined maximum attempts $T \in \{ 10,20,50\}$ for PC(1024-512), compared with CA-SCL decoding with list size $L \in \{2,4,8\}$.}\label{Fig.stateofart}
  \end{figure}
  
\begin{figure}[tb]
    \centering
    \includegraphics[width=8cm]{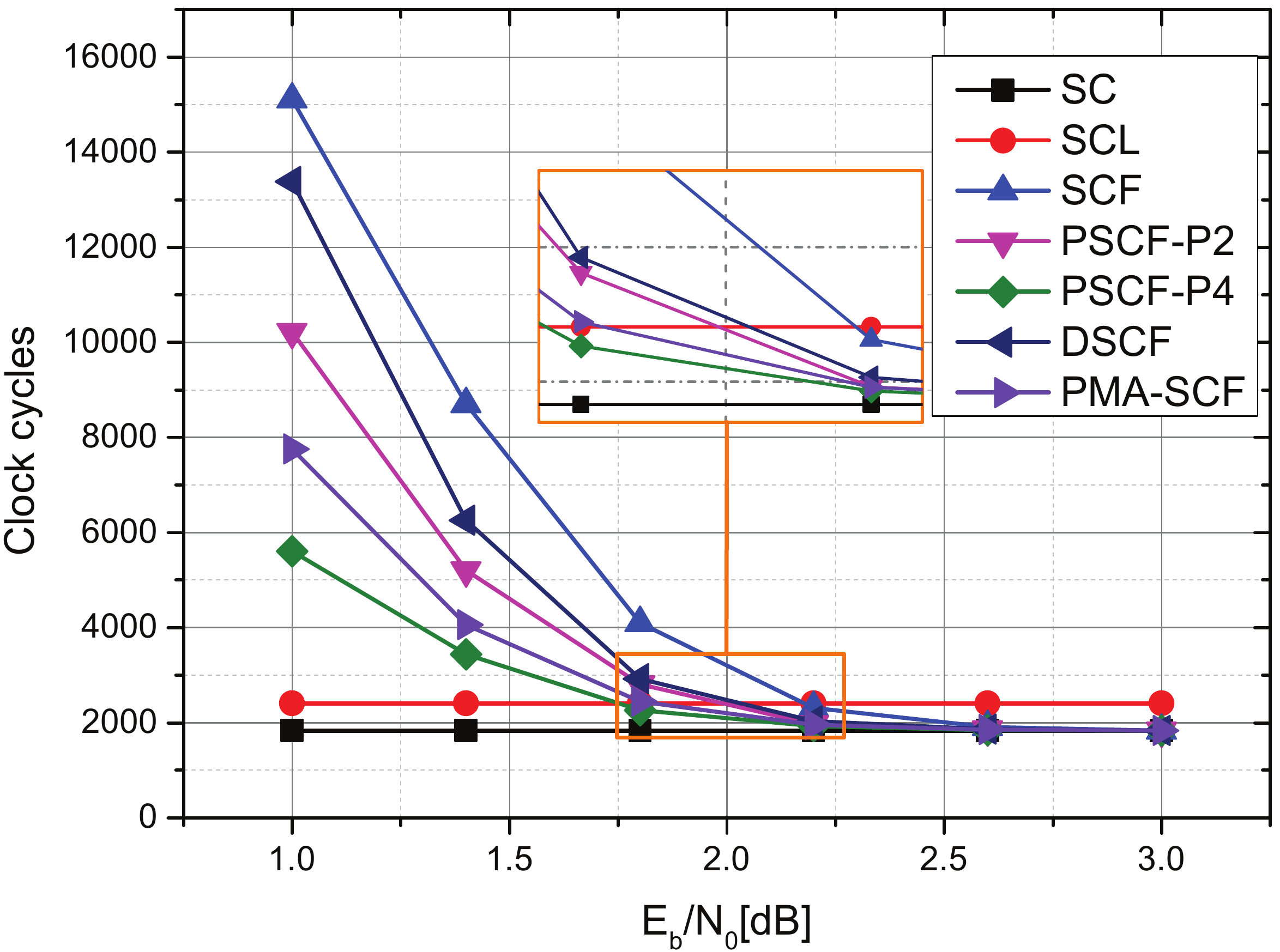}
    \caption{Average decoding latency of various decoding algorithm for $PC(1024,512)$. $T=20$ for all SCF based decoders.}\label{Fig.decode_latency}
  \end{figure}

   In order to evaluate the performance of PMA-SCF decoder with different predetermined maximum attempts $T$, they are compared with CRC-aided SCL decoding algorithm with list size $L \in \{2,4,8\}$ and 16-CRC 
   for PC(1024-528). As shown in Fig.\ref{Fig.stateofart}, the FER performance of oracle-assisted SC-oracle decoder with order 1 (SCO-O1) and
   order 2 (SCO-O2) are used as the baseline for comparison. One can observe that the FER performance of PMA-SCF decoding algorithm with different predetermined maximum attempts are all between the SCO-O1 and SCO-2 curves, while 
   only the CRC aided SCL decoding with list size $L=2$ (CA-SCL-L2) is worse than SCO-O1 and the CA-SCL-L8 is better than SCO-O2. The performance of PMA-SCF decoding with $T=20$ is similar to that of CA-SCL-L4 at almost all cases. Considering their decoding latency shown
   in Fig.\ref{Fig.decode_latency} and implementation complexity, the PMA-SCF decoding is more efficient than CA-SCL-L4 to get equivalent FER. 
   
   In Fig.\ref{Fig.decode_latency}, the decoding latency of our proposed decoding algorithm is evaluated, with respect to that of SCF, PSCF and DSCF.
   In this comparison, we use the clock cycles as the measurement, instead of the average number of attempts, since the adopting of pipeline architecture.
   The decoding latency of the SC decoder and the SCL decoder measured as that does in \cite{Giard2017PolarBear} are portrayed as the reference line. 
   The average decoding latency at each ${{{E_b}} \mathord{\left/
   {\vphantom {{{E_b}} {{N_0}}}} \right.
   \kern-\nulldelimiterspace} {{N_0}}}$ point is obtained by simulating $1 \times {10^8}$ frames.
   It can be observed that the decoding latency of SCF is the highest among all SCF based decoders at low ${{{E_b}} \mathord{\left/
   {\vphantom {{{E_b}} {{N_0}}}} \right.
   \kern-\nulldelimiterspace} {{N_0}}}$ regime, since the absolute value of LLR is not efficient to be used as the bit-flipping metric. Compared with
   SCF decoding, the PSCF decoding has much lower decoding latency, since it may stop decoding when a partition fails before $T$ iterations.
   Our proposed PMA-SCF is about 24\% above that of PSCF with $P=4$ at the worst ${{{E_b}} \mathord{\left/
   {\vphantom {{{E_b}} {{N_0}}}} \right.
   \kern-\nulldelimiterspace} {{N_0}}} = 1dB$, while it is up to ${\rm{1}}{\rm{.8}} \times$ faster than that of DSCF. 
   It can also be observed that the decoding latency of our algorithm degrades quickly as the ${{{E_b}} \mathord{\left/
   {\vphantom {{{E_b}} {{N_0}}}} \right.
   \kern-\nulldelimiterspace} {{N_0}}}$ increases, and approaches that of SCL at moderate ${{{E_b}} \mathord{\left/
   {\vphantom {{{E_b}} {{N_0}}}} \right.
   \kern-\nulldelimiterspace} {{N_0}}}$.
   At higher ${{{E_b}} \mathord{\left/
   {\vphantom {{{E_b}} {{N_0}}}} \right.
   \kern-\nulldelimiterspace} {{N_0}}}$, all SCF based decoding converge to the decoding latency of SC algorithm.
  
  \section{Conclusion}
  In this paper, we propose the PMA-SCF decoding algorithm, that generates the 
  bit-flipping list according to its bit-flipping metric and path metric, which provides an effective starting point to correct more erroneous decisions.
  The corresponding pipeline architecture is designed to reduce the decoding latency. We show that the average latency
  is much lower than current bit-flipping decoding method at the cost of increased memory.
  The simulation results show that our decoding algorithm can provide a performance improvement of up to 0.25dB at FER of $10^{-4}$ 
  compared to SCF decoding, while decode up to ${\rm{1}}{\rm{.8}} \times$ faster than DSCF decoding at ${{{E_b}} \mathord{\left/
  {\vphantom {{{E_b}} {{N_0}}}} \right.
  \kern-\nulldelimiterspace} {{N_0}}} = 1dB$ point. 

\bibliographystyle{IEEEtran}\bibliography{./PMA_SCF.bib}

\end{document}